\documentclass{article}
\usepackage{spconf,amsmath}
\usepackage{amsthm}
\usepackage{amsfonts}
\usepackage{amssymb}
\usepackage{mathrsfs}
\usepackage{mathtools}
\usepackage[english]{babel}
\usepackage{float}
\usepackage[toc,page]{appendix}
\usepackage[pdftex]{graphicx}
\usepackage{epstopdf}
\usepackage{amsmath}
\usepackage{color}
\usepackage{multirow}
\usepackage{graphicx}
\usepackage{IEEEtrantools}
\usepackage{bm}
\usepackage{balance}
\usepackage{microtype}
\usepackage[caption=false,font=footnotesize]{subfig}
\usepackage{cite}

\numberwithin{equation}{section}

\newcommand\nohyph{\hyphenpenalty=10000\relax\exhyphenpenalty=10000\relax} 

\newtheorem{theorem}{Theorem}[section]

\newtheorem{corollary}[theorem]{Corollary}

\newtheorem{proposition}[theorem]{Proposition}

\ninept
\usepackage{algorithm} 

\title{Performance of time delay estimation in a cognitive radar}
\name{Kumar Vijay Mishra and Yonina C. Eldar \thanks{\nohyph This work was funded by the European Union’s Horizon 2020 research and innovation programme under grant agreement ERC-BNYQ. K.V.M. acknowledges partial support via Andrew and Erna Finci Viterbi Fellowship.}}
\address{Andrew and Erna Viterbi Faculty of Electrical Engineering,\\ Technion -– Israel Institute of Technology, Haifa, Israel}
\begin{document}
\maketitle

\begin{abstract}
A cognitive radar adapts the transmit waveform in response to changes in the radar and target environment. In this work, we analyze the recently proposed sub-Nyquist cognitive radar wherein the total transmit power in a multi-band cognitive waveform remains the same as its full-band conventional counterpart. For such a system, we derive lower bounds on the mean-squared-error (MSE) of a single-target time delay estimate. We formulate a procedure to select the optimal bands, and recommend distribution of the total power in different bands to enhance the accuracy of delay estimation. In particular, using Cram\'{e}r-Rao bounds, we show that equi-width subbands in cognitive radar always have better delay estimation than the conventional radar. Further analysis using Ziv-Zakai bound reveals that cognitive radar performs well in low signal-to-noise (SNR) regions.
\end{abstract}


\maketitle

\section{Introduction}
\label{sec:intro}
In recent years, cognitive radar has garnered considerable research interest. The main advantage of such a system is its ability to learn the target environment and then adapt both the transmit and receive processing for an optimal performance \cite{haykin2006cognitive}. Conventional radars can also optimize and change their processing techniques depending on the target scene, but their adaptability is restricted to receive processing only. Several possible radar cognition capabilities have been suggested (see e.g. \cite{guerci2010cognitive,haykin2012cognitive}) where the environment specifications and corresponding suitable adaptive behaviors vary widely. Some common examples include introducing nulls towards the direction of clutter in the transmit antenna pattern \cite{guerci2010cognitive}, transmit beam-scheduling based on previous tracking history of the targets \cite{sharaga2015optimal,bell2015cognitive}, and designing transmit waveform code that avoids interfering bands by other licensed services \cite{aubry2015new,stinco2016spectrum}. In this work, we consider cognitive radar in the latter context where the transmit waveform is restricted to certain frequencies of interest. 

Many radar systems operate within a crowded electromagnetic spectrum where radio-frequency (RF) interference due to communication services is increasingly common \cite{griffiths2015radar}. In conventional radars, the RF interference is notched out at the receiver leading to problems of signal distortion (see e.g. \cite{davis2011foliage}). Some recent works, therefore, focus on designing radar waveforms such that the interference within the transmit signal spectrum is minimized \cite{he2010waveform,la2013design,aubry2015new}. While effective in mitigating RF interference, such a spectral shaping of transmit signal causes loss of signal bandwidth and radar resolution.

Recently, \cite{cohen2016towards} proposed a \textit{sub-Nyquist cognitive radar} (CR hereafter) where the transmit waveform is confined to a few disjoint subbands and the receiver samples and processes only these subbands. This system exploits Xampling \cite{mishali2011xampling} that has been used for reducing the sampling rate at the receiver  without sacrificing the range resolution. Since the total transmit power remains the same as the original full-band signal, the CR waveform has more in-band power resulting in an increase in the receiver's signal-to-noise ratio (SNR). In \cite{cohen2016towards}, this benefit was demonstrated through software and hardware experiments. The CR has also been realized in a hardware prototype for a sub-Nyquist multiple-input multiple-output (MIMO) radar \cite{mishra2016cognitive}. 

Prior works on CR do not study the system performance in terms of estimation accuracy of target parameters, such as location and velocity. For a single-antenna radar, only one dimension (range) of the target location can be estimated. Since range is linearly proportional to time delay, it suffices to consider only the time delay estimation for single antenna radar. In this paper, we analyze the performance of CR for estimating the time delay for a single target. We first devise a procedure to select the subbands that have the minimum spectral interference; here, the existing spectral power from other services at each frequency, i.e., the radar environment map (REM) is considered known. Once the subbands are selected for CR, we derive lower bounds on mean-squared-error (MSE) in order to recommend a way to redistribute the total power among chosen bands so that the accuracy in time delay estimation is enhanced. Our work has several advantages: the CR transmission minimizes the spectral interference, delay estimation accuracy is enhanced by optimal distribution of power, the range resolution is not lost while sampling at a low rate, and a small portion of the available bandwidth is used.

Lower bounds on estimation errors are useful for evaluating performance \cite{ben2009lower}. A common bound is the Cram\'{e}r-Rao Lower Bound (CRLB) on the variance of the estimator of an unknown target parameter \cite{van2013detection}. The CRLB is a tight bound only when the errors are small \cite{zeira1993realizable}. It is well known \cite{weiss1983fundamentalnarrow,weinstein1984fundamentalwide} that the performance of the time delay estimators is characterized by the presence of distinct SNR regions. When the SNR is high (\textit{asymptotic} region), the CRLB describes the MSE accurately. In low SNR regions, the information from signal observations is insufficient and the MSE is close to that obtained via only \textit{a priori} information. In between these two limiting cases lies the transition region where the signal observations are subjected to ambiguities that do not factor in the CRLB \cite{van2013detection}. Other bounds that are tighter than the CRLB have been developed (see e.g. \cite{sadler2006survey} for a survey). In particular, the Ziv-Zakai lower bound (ZZLB) \cite{ziv1969some} can accurately identify the SNR thresholds that define the ambiguity region \cite{ianniello1982time}. 

In the context of radar signals, some examples of the CRLB for estimating time delay, velocity, and direction of arrival of the target are derived in, e.g., \cite{kay1998fundamentalsEst,peebles}. The CRLBs for time delay estimation have also been studied, jointly with Doppler, for active arrays \cite{dogandzic2001cramer}, MIMO \cite{he2010noncoherent}, and extended targets \cite{zhao64cramer}. The ZZLB treats the time delay as a parameter that is distributed uniformly at random over an interval. The ZZLBs for target location estimation have been derived for a conventional radar \cite{bell2015cognitive} and several different configurations of MIMO systems \cite{chiriac2010ziv,chiriac2015ziv}. In this work, we derive the CRLB and the ZZLB for estimating the time-delay for a single target in a cognitive radar where the spectrum is multi-band or \textit{dispersed}. Some previous works for time-delay estimation in communications have derived CRLB for a dispersed signal \cite{kocak2010time} without any maximum power constraint. 
We show that the delay estimation with cognitive radar is always more accurate than the conventional radar if all subbands have equal bandwidth. A comparison of ZZLB shows that cognitive radar delay estimates are more accurate in low SNR regions than the conventional radar even if their respective CRLBs are equal.

Our system model for CR is described in the next section. In Section \ref{sec:optbands}, we formulate a procedure to choose CR subbands when the REM is known. Section \ref{sec:crlb} presents the derivation of the CRLB for delay estimation with a CR and conditions for enhanced performance. We revisit the same problem for the derivation of ZZLB in Section \ref{sec:zzb}. We conclude with remarks and future work in Section \ref{sec:summ}. 

\section{System Model}
\label{sec:problem_formulation}
We consider a single-antenna radar system that transmits a pulse $h(t)$ towards the targets-of-interest. We assume that the transmit pulse is nonzero over the interval $[0, \tau]$. The transmit pulse spectrum is given by the continuous-time Fourier Transform (CTFT) of $h(t)$:\par\noindent\small
\begin{align}
H(\omega) = \int\limits_{-\infty}^{\infty} h(t)e^{-j\omega t}\, \mathrm{d} t \approx \int\limits_{-B_h/2}^{B_h/2} h(t)e^{-j\omega t}\, \mathrm{d} t,
\end{align}\normalsize
where the signal energy outside the bandwidth $B_h$ is negligible. Suppose a Swerling-1 target is located at range $R$ with respect to the radar, so that the round-trip time delay for the transmit pulse to the target and back is given by $\tau_0 = 2R/c_0$, where $c_0$ is the speed of light. We assume that the target environment is free of clutter and, except the range, all other target parameters such as its reflectivity and Doppler velocity are known. The continuous-time received signal at baseband is then\par\noindent\small
\begin{align}
\label{eq:rxsig_con}
x_R(t) = h(t-\tau_0) + \eta(t), \: 0\le t \le T_s,
\end{align}\normalsize
where $T_s$ is the observation interval such that $\tau_0 \in [0, T_s]$, and $\eta(t)$ is additive white Gaussian noise that is band-limited to $\mathcal{B}$ with two-sided power spectral density $\mathcal{N}_0/2$, and variance $\sigma^2 = \mathcal{N}_0B_h$. 

\begin{figure}[!t]
  \includegraphics[scale=0.25]{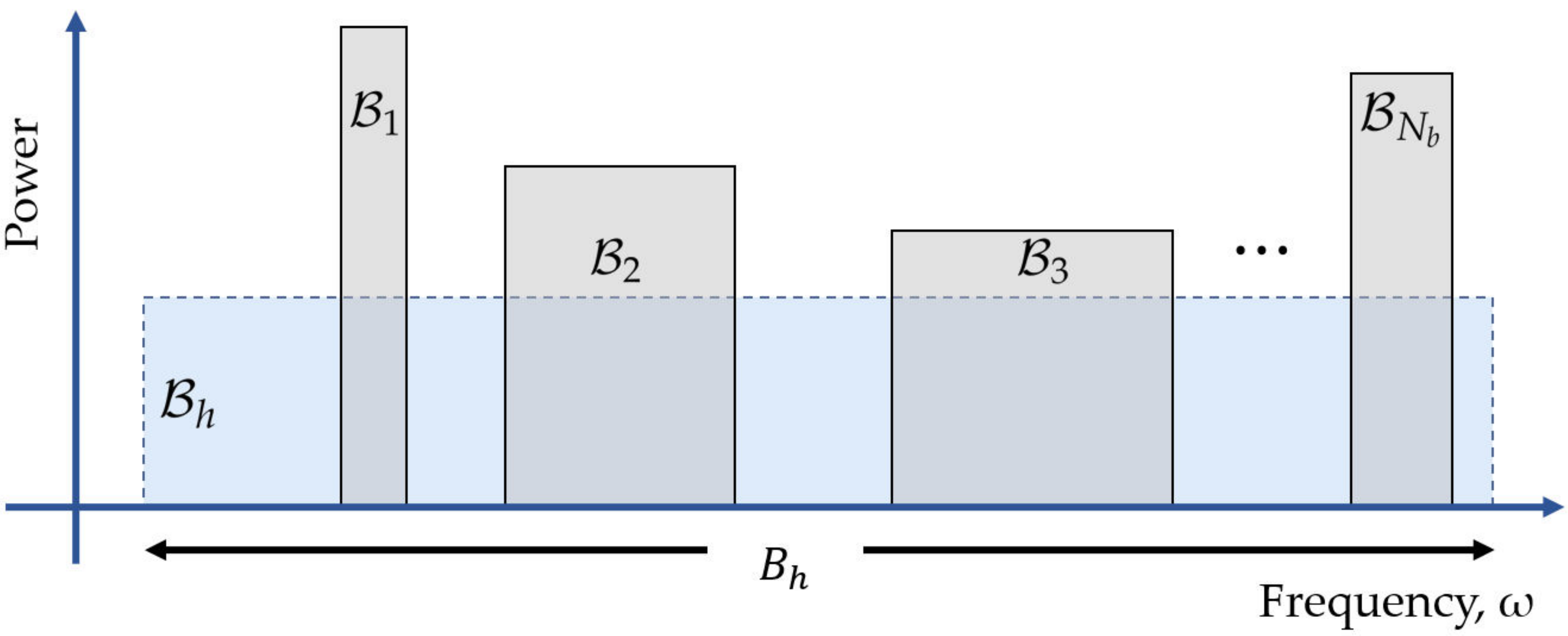}
  \caption{\footnotesize{A conventional radar with bandwidth $B_h$ transmits in the  band $\mathcal{B}_h$. A cognitive radar transmits only in subbands $\{\mathcal{B}_i\}_{i=1}^{N_b}$, but with more in-band power than the conventional radar. Only one-sided spectrum is shown here.}\vspace{-12pt}}
	\label{fig:cogspec}
\end{figure}

In our proposed CR, the spectrum of the transmit waveform $\widetilde{h}(t)$ is restricted to a total of $N_b$ non-contiguous frequency bands, each represented by the set $\mathcal{B}_i$, $1 \le i \le N_b$, of its constituent frequencies (Figure \ref{fig:cogspec}). We denote the center-frequency of the band $\mathcal{B}_i$ by $\omega_{C_i}$ and its bandwidth by $B_i$ so that $\mathcal{B}_i = [\omega_{C_i}-B_i/2, \omega_{C_i}+B_i/2]$. The CTFT of $\widetilde{h}(t)$ is given by\par\noindent\small
\begin{align}
\widetilde{H}(\omega) &= \begin{dcases} 
    \beta_iH(\omega),\: \omega \in \bigcup_{i=1}^{N_b}\left[\omega_{C_i}-\dfrac{B_i}{2}, \omega_{C_i}+\dfrac{B_i}{2}\right] \subset \mathcal{B}_h\\
    0, \:\text{otherwise},
   \end{dcases}
\end{align}\normalsize
where $\beta_i$'s are the \textit{power redistribution constants} that determine the proportion of total power $P$ to be distributed among different subbands. Importantly, the total transmit power $P$ of the CR remains the same such that the power relation between the conventional and cognitive waveforms:\par\noindent\small
\begin{align}
\int_{B_h} |H(\omega)|^2\, \mathrm{d}\omega = \sum_{i=1}^{N_b}\int_{B_i} |\widetilde{H}(\omega)|^2\, \mathrm{d}\omega = \sum\limits_{i=1}^{N_b}P_i = P,
\end{align}\normalsize
where the subband powers $P_i$ are higher than the full-band power.

The CR waveform consists of multiple subband waveforms $\widetilde{h}_i(t)$ combined together and transmitted over a single antenna. At the receiver, the signal is filtered into the respective subbands and then processed together. The received signal corresponding to the $i$th subband can be expressed as\par\noindent\small
\begin{align}
\label{eq:rxsig_cogi}
x_{CR_i}(t) = \widetilde{h_i}(t-\tau_0) + \eta_i(t), \: 0\le t \le T_s,\: 1 \le i \le N_b,
\end{align}\normalsize
where $\eta_i(t)$ is white Gaussian noise band-limited to $\mathcal{B}_i$, independent for different subbands, and with variance $\sigma_i^2 = \mathcal{N}_0B_i$. Note that while the signal power is enhanced in subbands, the noise spectral density remains the same. The spectral interference is considered negligible in the selected cognitive radar bands. By superimposing all the signals in (\ref{eq:rxsig_cogi}), we obtain the continuous-time CR received signal at baseband\par\noindent\small
\begin{align}
\label{eq:rxsig_cog}
x_{CR}(t) = \sum\limits_{i=1}^{N_b} x_{CR_{i}}(t) = \widetilde{h}(t-\tau_0) + \eta(t), \: 0\le t \le T_s.
\end{align}\normalsize

We define the signal-to-noise-ratio (SNR) for the signals $h(t)$, $\widetilde{h}_i(t)$, and $\widetilde{h}(t)$ respectively as,\par\noindent\small
\begin{align}
\textrm{SNR} = \frac{P}{\sigma^2} = \frac{P}{\mathcal{N}_0B_h};\:\textrm{SNR}_i = \frac{P_i}{\sigma_i^2}=\frac{P}{\mathcal{N}_0B_i};\nonumber\\
\widetilde{\textrm{SNR}} = \left.{\sum\limits_{i=1}^{N_b}P_i} \middle/ {\sum\limits_{i=1}^{N_b}\sigma_i^2} \right. = \left. {\sum\limits_{i=1}^{N_b}P_i} \middle/ \left({\mathcal{N}_0\sum\limits_{i=1}^{N_b}B_i}\right)\right. .
\end{align}\normalsize
We note that $\widetilde{SNR}>SNR$. For the low-pass signal $h(t)$, we define its root-mean-square (rms) bandwidth $\overline{F}$ as\par\noindent\small
\begin{align}
\overline{F} = \sqrt{\dfrac{\int\limits_{-B_h}^{B_h} |H(\omega)|^2 \omega^2\, \mathrm{d}\omega}{\int\limits_{-B_h}^{B_h} |H(\omega)|^2\, \mathrm{d}\omega}} = \sqrt{\dfrac{\int\limits_{0}^{T_s} [h'(t)]^2\, \mathrm{d}t}{\int\limits_{0}^{T_s} h^2(t)\, \mathrm{d}t}},
\end{align}\normalsize
where we define $h'(t)$ as the first derivative of $h(t)$. For the second derivative, we use the notation $h''(t)$. The rms bandwidth for the band-pass signal $\widetilde{h}_i(t)$ is, \par\noindent\small
\begin{align}
\overline{F_i} = 2\sqrt{\dfrac{\int\limits_{\omega_{C_i}-B_i/2}^{\omega_{C_i}+B_i/2} |\widetilde{H}_i(\omega)|^2 (\omega-\omega_{C_i})^2\, \mathrm{d}\omega}{\int\limits_{\omega_{C_i}-B_i/2}^{\omega_{C_i}+B_i/2} |\widetilde{H}_i(\omega)|^2\, \mathrm{d}\omega}} = \sqrt{\dfrac{\int_{0}^{T_s} [\widetilde{h_i}'(t)]^2 \, \mathrm{d}t}{\int_{0}^{T_s} \widetilde{h_i}^2(t)\, \mathrm{d}t}},
\end{align}\normalsize
where a factor of $2$ in the first equality occurs because the integrals inside the square-root measure the rms bandwidth only on one side of the spectrum. The rms bandwidth measures the ``width'' or the variance of the power spectral density, and is smaller than or equal to the full bandwidth. We can write for some constants $\alpha$ and $\alpha_i$'s:\par\noindent\small
\begin{align}
\overline{F} = \alpha B_h,\:\alpha\le 1;\:\overline{F_i} = \alpha_i B_i,\:\alpha_i\le 1.
\end{align}\normalsize
For example, if the signals have flat CTFTs, then $\alpha = 1/\sqrt{3}$.

\section{Selection of Subbands}
\label{sec:optbands}
A major application of our cognitive radar is to avoid the spectral interference from other systems. In this section, we develop a procedure to select the subbands in the set $\mathcal{B}_h$ such that the subbands have the least RF interference. The maximum number of CR subbands can be fixed to $N_b$ due to the receiver hardware constraints. We assume the REM -- a spectral map in the form of interfering energy levels at all frequencies -- is known to the radar user. We choose a Discrete Frequency Transform (DFT) grid of $M$ points which is large enough so that it densely covers the set $\mathcal{B}_h$. Suppose the measurements of REM are available in the form of an $M$-point frequency-domain vector $\mathbf{y}$. The information available from the REM may also indicate that the CR waveform must avoid all the frequencies in a certain set $\mathcal{B}_C \subset \mathcal{B}_h$, which is decidedly used by another service. We seek an $M$-point frequency response vector $\mathbf{z}$ that is nearly the same as the spectrum of $\mathbf{y}$ in a total of $N_b$ lowest interference subbands and vanishes otherwise. We formulate $\mathbf{z}$ as an $N_b$-block-sparse vector that consists of segments $\mathbf{z}_1$, $\mathbf{z}_2$, $\cdots$, $\mathbf{z}_{N_b}$ of lengths $d_1$, $d_2$, $\cdots$, $d_{N_b}$, respectively. It can be obtained by solving the optimization problem\par\noindent\small
\begin{align}
	\label{eq:band_select1}
	& \underset{\mathbf{z}}{\text{minimize}} \; (\left\Vert\mathbf{z}_1\right\Vert_2 + \left\Vert\mathbf{z}_2\right\Vert_2 + \cdots + \left\Vert\mathbf{z}_{N_b}\right\Vert_{2}) \nonumber\\
	& \;\;\; \text{subject to}  \;\; \left\Vert\mathbf{z}_{\mathcal{B}_h\setminus\mathcal{B}_C} - \mathbf{y}_{\mathcal{B}_h\setminus\mathcal{B}_C}\right\Vert_2 \le \xi\nonumber\\
    & \quad\quad\quad\quad \;\;\;\; \mathbf{z}_{\mathcal{B}_C} = 0 
\end{align}\normalsize
where $\xi$ is a positive constant. When the width of blocks in $\mathbf{Z}$ are known (due to receiver hardware design constraints), but the locations of the blocks are unknown, then greedy algorithms such as StructOMP \cite{huang2011learning} can be employed to solve this optimization problem. In the above, additional constraints of minimum bandwidths, constant power across all bands, range sidelobe levels, and minimum separation between the bands may also be imposed. For example, the hardware design constraints may require all bands to be equal for simplicity, i.e., $d_1 = d_2 = \cdots = d_{N_b}$. 

The optimal subbands are the support of the vector $\mathbf{z}$. Once the CR subbands are identified, we would like to know an optimal way to distribute the total power in different subbands. We address this question in the context of the accuracy of time delay estimation in the next section.

\section{CRLB for Delay Estimation} 
\label{sec:crlb}
Our goal is to develop and compare the CRLB expressions for estimating time delays from the received signal in conventional (\ref{eq:rxsig_con}) and cognitive radars (\ref{eq:rxsig_cog}). We then use these bounds to determine the choice of $\beta_i$ so that the accuracy is higher in CR.

The real-valued deterministic signal $h(t)$ in (\ref{eq:rxsig_con}) and $\widetilde{h}(t)$ in (\ref{eq:rxsig_cog}) are known except for the parameter $\tau_0$. The deterministic CRLB for the estimate $\hat{\tau}_0$ of $\tau_0$ in case of the conventional radar is well-known to be \cite{peebles,helstrom}\par\noindent\small
\begin{align}
\label{eq:crlb_con}
CRLB_R(\hat{\tau}_0) = \dfrac{\sigma^2}{P \overline{F^2}} = \dfrac{1}{SNR\cdot\overline{F^2}}.
\end{align}\normalsize
In case of CR, we have the following result.
\begin{theorem}
\label{thm:crlb_cog}
The CRLB for estimating delay with the cognitive radar received signal (\ref{eq:rxsig_cog}) is given by\par\noindent\small
\begin{align}
\label{eq:crlb_cog}
CRLB_{CR}(\hat{\tau}_0) = \dfrac{1}{\sum\limits_{i=1}^{N_b}SNR_i\cdot\overline{F_i^2}}.
\end{align}\normalsize
\end{theorem}
\begin{IEEEproof} Let $\theta = \tau_0$ be the unknown signal parameter. If all the subband signals in (\ref{eq:rxsig_cogi}) are observed over the duration $T_s$, then the log-likelihood function of $\theta$ is\par\noindent\small
\begin{align}
\label{eq:llf_cog}
\mathcal{L}(\theta) = c - \sum\limits_{i=1}^{N_b}\dfrac{1}{2\sigma^2_i}\int\limits_0^{T_s}|x_{CR_{i}}(t) - \widetilde{h_i}(t-\theta)|^2\, \mathrm{d}t,
\end{align}\normalsize
where $c$ is a constant independent of the unknown parameter $\theta$. Differentiating the log-likelihood once produces\par\noindent\small
\begin{align}
\frac{\partial \mathcal{L}(\theta)}{\partial \theta} = \sum\limits_{i=1}^{N_b}\dfrac{1}{\sigma^2_i}\int\limits_0^{T_s}(x_{CR_{i}}(t) - \widetilde{h_i}(t-\theta)) \widetilde{h_i}'(t-\theta)\, \mathrm{d}t,
\end{align}\normalsize
and a second differentiation results in\par\noindent\small
\begin{flalign}
\frac{\partial^2 \mathcal{L}(\theta)}{\partial \theta^2} &= \sum\limits_{i=1}^{N_b}\dfrac{1}{\sigma^2_i}\int\limits_0^{T_s}\{(x_{CR_{i}}(t) - \widetilde{h_i}(t-\theta)) \widetilde{h_i}''(t-\theta) \nonumber\\
&\:\:\:\:\:\:\:\:\:\:\:\:\:\:\:\:\:\:\:\:\:\:\:\:\:\:\:\:- [\widetilde{h_i}'(t-\theta)]^2\}\, \mathrm{d}t.
\end{flalign}\normalsize
Taking the negative of the expected value yields the Fisher information $I(\theta)$ for the observed data\par\noindent\small
\begin{align}
I(\theta) &= -E\left[\frac{\partial^2\mathcal{L}(\theta)}{\partial\theta^2}\right] = \sum\limits_{i=1}^{N_b} \frac{1}{\sigma_i^2} \int_{0}^{T_s} [\widetilde{h_i}'(t-\theta)]^2 \, \mathrm{d}t\nonumber\\
&=\sum\limits_{i=1}^{N_b} \frac{1}{\sigma_i^2} \int_{0}^{T_s} [\widetilde{h_i}'(t)]^2 \, \mathrm{d}t,
\end{align}\normalsize
where we note that the expectation is taken with respect to the pdf of $x_{CR_{i}}$ as a function of $\theta$. The last equality is obtained by changing the variables and noting that the signal is nonzero only over the interval $\tau_0 \le t \le \tau_0+T_s$. The CRLB is then\par\noindent\small
\begin{align}
CRLB_{CR}(\hat{\tau}_0) &= \dfrac{1}{I(\theta)} = \dfrac{1}{\sum\limits_{i=1}^{N_b} \dfrac{P_i}{\sigma_i^2} \dfrac{\int_{0}^{T_s} [\widetilde{h_i}'(t)]^2 \, \mathrm{d}t}{\int_{0}^{T_s} \widetilde{h_i}^2(t)\, \mathrm{d}t}}= \dfrac{1}{\sum\limits_{i=1}^{N_b} SNR_i\cdot\overline{F_i^2}},
\end{align}\normalsize
completing the proof.
\end{IEEEproof}
By comparing the CRLBs of conventional and cognitive radars, we can find the distribution of the total power in different subbands for better performance of cognitive radar.
\begin{proposition}
\label{prop:crlb_cog_con}
Given the subbands $B_i$ and the full-band $B_h$, the delay estimation with CR is not worse than the conventional radar if the in-band powers obey the following relation\par\noindent\small
\begin{align}
\label{eq:power_relation1}
\sum\limits_{i=1}^{N_b} P_i \alpha_i^2 B_i &\ge P\alpha^2 B_h.
\end{align}\normalsize
\end{proposition}
\begin{IEEEproof} 
The $CRLB_{CR}(\hat{\tau_0})$ is lower than $CRLB_{R}(\hat{\tau_0})$, if \par\noindent\small
\begin{align}
\sum\limits_{i=1}^{N_b} \frac{P_i}{\sigma^2_i}\cdot\overline{F_i^2} &\ge \frac{P}{\sigma^2}\cdot\overline{F^2},\nonumber
\end{align}\normalsize
or\par\noindent\small
\begin{align}
\sum\limits_{i=1}^{N_b} \frac{P_i}{\mathcal{N}_0 B_i}\cdot\alpha_i^2 B_i^2 &\ge \frac{P}{\mathcal{N}_0 B_h}\cdot\alpha^2 B_h^2\nonumber,
\end{align}\normalsize
from where the inequality in the proposition follows.
\end{IEEEproof}
\begin{corollary}
\label{cor:crlb_flat_psd}
Suppose the signals in (\ref{eq:rxsig_con}) and (\ref{eq:rxsig_cogi}) have flat CTFTs over their respective bandwidths $B_h$ and $B_i$. Then the condition in Proposition \ref{prop:crlb_cog_con} reduces to\par\noindent\small
\begin{align}
\sum\limits_{i=1}^{N_b} \beta^2_iB_i^2 &\ge B_h^2.
\end{align}\normalsize
\end{corollary}
\begin{IEEEproof}
Let the CTFTs have a constant value of $A$. Then $P = A^2B_h$ and $P_i = A^2\beta^2_iB_i$. Also, $\alpha = \alpha_i = \cdots = \alpha_{N_b} = 1/\sqrt{3}$. Substituting these in (\ref{eq:power_relation1}) completes the proof.
\end{IEEEproof}
\begin{corollary}
\label{clm:crlb_equiwidth}
In Corollary \ref{cor:crlb_flat_psd}, if all CR subbands have same bandwidth, then (\ref{eq:power_relation1}) always holds true.
\end{corollary}
\begin{IEEEproof}
In (\ref{eq:power_relation1}), $B_i$s can be written as some fraction $\gamma$ of $B_h$ so that $P = \sum\limits_{i=1}^{N_b} P_i \ge \gamma P$, which is always true.
\end{IEEEproof}
\begin{corollary}
\label{clm:crlb_equiheight}
In Corollary \ref{cor:crlb_flat_psd}, if all the CR subbands have same CTFT value, i.e. $\beta_1 = \cdots = \beta_{N_b} = \beta$, then $\beta \ge B_h/({\sum\limits_{i=1}^{N_b}B_i^2})^{\frac{1}{1}}$. 
\end{corollary}
Since the CRLB analyses describes MSE only at high SNRs, we turn to ZZLB comparison now for low SNR situations.
\section{ZZLB for delay estimation} 
\label{sec:zzb}
In order to derive the ZZLB for the signal model as in (\ref{eq:rxsig_con}), we assume $\tau_0$ is distributed uniformly at random over the interval $[0, T_s]$, with the pdf $p_{\tau_0}(\tau_0)$ and prior variance $\sigma^2_{\tau_0}$. For the scalar parameters, the ZZLB bounds for the mean squared error (MSE)\par\noindent\small
\begin{align}
\overline{\epsilon}^2 = E\{\epsilon^2\} = E\{|\hat{\tau_0} - \tau_0|^2\},
\end{align}\normalsize
by averaging all errors $\delta$ over their probabilities $\textrm{Pr}(|\epsilon| \ge \delta/2)$ using the following identity \cite{bell1997extended}\par\noindent\small
\begin{align}
\overline{\epsilon}^2 = \dfrac{1}{2}\int\limits_{0}^{\infty}\textrm{Pr}\left(|\epsilon| \ge \delta/2\right)\delta\, \mathrm{d}\delta.
\end{align}\normalsize
A detection theoretic approach for a binary hypotheses problem reveals that $\textrm{Pr}(|\epsilon| \ge \delta/2)$ is the probability of $\tau_0$ taking a value $\widetilde{\tau_0}$ ($\mathcal{H}_0$ or null hypothesis) or $\widetilde{\tau_0}+\delta$ ($\mathcal{H}_1$ or alternative hypothesis). The basic ZZLB expression is the following integral \cite{chazan1975improved}\par\noindent\small
\begin{align}
\overline{\epsilon}^2 \ge \frac{1}{T_s}\int\limits_{0}^{T_s}\delta\, \mathrm{d}\delta \int\limits_{0}^{T_s-\delta}P_e(\widetilde{\tau}_0, \widetilde{\tau}_0 + \delta)\, \mathrm{d}\widetilde{\tau_0},
\end{align}\normalsize
where $P_e(\widetilde{\tau}_0, \widetilde{\tau}_0 + \delta)$ is the minimum attainable probability of error in the likelihood ratio test between the two hypotheses. Typically, the ZZLB results in complex and intractable integrals. 

For delay estimation in a conventional single-antenna radar, with signal model as in (\ref{eq:rxsig_con}), \cite{bell1997extended} derived a closed-form expression for a weaker version of the so-called extended ZZLB (EZB). The key assumption made here is that the transmit pulse-width $\tau$ is short compared to the interval $T_s$. This assumption simplifies the EZB for the conventional radar as \cite{bell1997extended}\par\noindent\small
\begin{align}
\label{eq:ezzlb_con}
EZB_R(\hat{\tau_0}) = \sigma^2_{\tau_0}\cdot 2Q\left(\sqrt{\dfrac{SNR}{2}}\right) + \dfrac{\Gamma_{3/2}\left(\dfrac{SNR}{4}\right)}{SNR\cdot\overline{F}^2},
\end{align}\normalsize
where $Q(\cdot)$ denotes the right tail Gaussian probability function and $\Gamma_{a}(b)$ is the incomplete gamma function with parameter $a$ and upper limit $b$. The behavior of the EZB in (\ref{eq:ezzlb_con}) can be summarized as follows. When the SNR $\rightarrow 0$, then the right-tail probability function approaches unity and the incomplete gamma function vanishes; therefore, in low SNR situations, the EZB is determined mostly by the \textit{a priori} distribution through $\sigma^2_{\tau_0}$. However, when SNR $\rightarrow \infty$, then $Q(\cdot)$ vanishes and $\Gamma_{3/2}(\cdot)$ approaches unity. Therefore, the high-SNR EZB converges to $1/(SNR\cdot\overline{F^2})$, which is same as the CRLB obtained in (\ref{eq:crlb_con}). The exact thresholds for the asymptotic and \textit{a priori} regions can be found using techniques from \cite{weiss1983fundamentalnarrow,weinstein1984fundamentalwide,bell1997extended}.

To derive an expression for cognitive radar similar to the EZB, we make the same assumption that the transmit pulse of the cognitive radar is short when compared to $T_s$. 
\begin{theorem}
\label{thm:ezzlb_cog}
The EZB for estimating delay with the cognitive radar received signal (\ref{eq:rxsig_cog}) is given by\par\noindent\small
\begin{flalign}
\label{eq:ezzlb_cog}
EZB_{CR}(\hat{\tau}_0) &= \sigma^2_{\tau_0}\cdot 2Q\left(\sqrt{\dfrac{\widetilde{SNR}} {2}}\right) + \dfrac{\Gamma_{3/2}\left(\dfrac{\widetilde{SNR}}{4}\right)}{\sum\limits_{i=1}^{N_b}SNR_i\cdot\overline{F_i^2}}.
\end{flalign}\normalsize
\end{theorem}
We omit the detailed derivation of the result in Theorem \ref{thm:ezzlb_cog} due to space. Suppose the in-band powers are distributed such that the CRLBs of both conventional and cognitive radars are same (i.e. $SNR\cdot\overline{F^2} = \sum\limits_{i=1}^{N_b}SNR_i\cdot\overline{F_i^2}$). Then, their EZB comparison provides an important revelation: since the $\widetilde{SNR} > SNR$ for a given power $P$, the SNR threshold for asymptotic performance of $EZB_{CR}$ is lower than $EZB_{R}$. In other words, as the noise increases and power remains constant for both radars, the asymptotic performance of $EZB_{CR}$ is more tolerant to the noise than the $EZB_{R}$.

\section{Summary}
\label{sec:summ}
We devised procedures to identify appropriate subbands and distribution of in-band powers for a single-antenna CR, where we derived the CRLB and ZZLB for time-delay estimation of a single target. The CRLB comparison of cognitive and conventional radar provides conditions for distributing the total power in different subbands. We restricted ourselves to a single target and single pulse. If multiple pulses are being transmitted, then the energy of the signal is enhanced by a factor given by the number of pulses and the analysis is analogous. The multiple point targets merit consideration of the range sidelobes. In future work, it would be interesting to provide CR power conditions for accuracy of estimating other parameters such as Doppler velocity. 

\clearpage 
\balance
\bibliographystyle{IEEEtran}
\bibliography{main}

\end{document}